\documentclass[12pt]{article}
\pdfoutput=1
\usepackage{jhep-mod}
\usepackage{bm}
\usepackage{amssymb}% http://ctan.org/pkg/amssymb
\usepackage{pifont}% http://ctan.org/pkg/pifont
\usepackage{slashed} % for Dirac operator
\usepackage{slantsc} % slanted smallcaps...
\def\d{{\mathrm{d}}}

\title{Global properties of physically interesting Lorentzian spacetimes}
\author{Deloshan Nawarajan\,\hbox{{\sf and}}}
\emailAdd{\\ deloshan.nawarajan@msor.vuw.ac.nz}
\author{Matt Visser\,}
\emailAdd{matt.visser@msor.vuw.ac.nz}
\affiliation{ \mbox{School of Mathematics and Statistics,}\\
Victoria University of Wellington; \\
PO Box 600, Wellington 6140, New Zealand.}
\abstract{\\
Under normal circumstances most members of the general relativity community focus almost exclusively on the  local properties of spacetime, such as the locally Euclidean structure of the manifold and the Lorentzian signature of the metric tensor. 
When combined with the classical Einstein field equations this gives an extremely successful empirical model of classical gravity and classical matter --- at least as long as one does not ask too many awkward questions about global issues, (such as global topology and global causal structure).  
We feel however that this is a tactical error --- even without invoking full-fledged ``quantum gravity'' we know that the standard model of particle physics is also an extremely  good representation of some parts of empirical reality; 
and we had better be able to carry over all the good features of the standard model of particle physics --- at least into the realm of semi-classical quantum gravity. 
Doing so gives us some interesting global features that spacetime should possess: On physical grounds spacetime should be space-orientable, time-orientable, and spacetime-orientable, and it should possess a globally defined tetrad (vierbein, or in general a globally defined vielbein/$n$-bein). So on physical grounds spacetime should be parallelizable. This strongly suggests that the metric is not the fundamental physical quantity; a very good case can be made for the tetrad being more fundamental than the metric.  
Furthermore, a globally-defined ``almost complex structure'' is almost unavoidable.
Ideas along these lines have previously been mooted, but much is buried in the pre-{\sf arXiv} literature and is either forgotten or inaccessible. We shall revisit these ideas taking a perspective very much based on empirical physical observation.

\medskip
\noindent
D{\sc{ate}}: 14 January 2016; \LaTeX-ed \today.
}
\begin{document}
%------------------------------------------------------------------------------------------------------------------------------------------
\maketitle
%------------------------------------------------------------------------------------------------------------------------------------------
\parskip 5pt
%------------------------------------------------------------------------------------------------------------------------------------------
%------------------------------------------------------------------------------------------------------------------------------------------
%------------------------------------------------------------------------------------------------------------------------------------------
\def\R{{\mathbb{R}}}
\def\Q{{\mathbb{Q}}}
\def\C{{\mathbb{C}}}
%------------------------------------------------------------------------------------------------------------------------------------------
\clearpage
%%----------------------------------------------------------------------------------------------------------------------------------------
\section{Introduction}
%%----------------------------------------------------------------------------------------------------------------------------------------
Standard general relativity makes only quasi-local statements about the behaviour of finite regions of spacetime; typically only within a single coordinate patch.
That is, standard general relativity, (at least as presented in most textbooks and introductory courses), is really a quasi-local theory. (As opposed to being ultra-local, with physics defined only at a point, or global, with physics completely under control \emph{everywhere} in the universe.)~\footnote{We shall eschew the use of the word ``local'', on the grounds that it has by now been so grossly misused as to become almost meaningless, in favour of ``quasi-local'' (patch-wise physics) or ``ultra-local'' (point-wise physics).} 

On the one hand this quasi-local focus is entirely reasonable,  since laboratory and observational physicists can only measure/observe finite regions over finite time intervals. (So for instance apparent/trapping horizons are good physics, whereas event horizons are [in a precise technical sense] not physically observable~\cite{observability}.) 
The basic reason for this quasi-local limitation is that standard general relativity focusses almost exclusively on the local differential properties of the metric tensor, and its interaction with the classical stress-energy tensor (often modelled by dust, perfect fluid, or the like). 

On the other hand, the standard model of particle physics contains qualitatively different objects, (such as chiral fermions,  and the  $C$, $P$, and $T$ operators, \emph{etcetera}), all of which are observed to still work rather well in weak gravitational fields~\cite{Hollands:2004, Hollands:2008, Hollands:2014}. 
This very strongly indicates that we should, (at the very least), bootstrap the standard model of particle physics into the realm of semi-classical gravity. Doing so requires us to make some choices, tantamount to making some global choices regarding the global structure of spacetime~\cite{Lee1, Lee2, Geroch}. 
(Some of these ideas have been mooted previously, but have often fallen behind the {\sf arXiv}-horizon, or even the {\sf Google}-horizon, and subsequently been forgotten. We shall revisit these ideas taking a very physical and empirically based viewpoint.)

We shall explore some of the inter-related issues arising from these considerations, focussing largely on space-orientability, time-orientability, and spacetime-orientability, and the existence of globally defined tetrads (vierbeins, or in general dimensionality,  globally defined vielbeins/$n$-beins), otherwise known as ``frame fields''. The existence of globally defined tetrads, either (time)+(space) tetrads, complex null tetrads, or more complicated linearly independent tetrads, is equivalent to the assertion  that spacetime is parallelizable. 

Furthermore, a globally-defined ``almost complex structure'' is almost unavoidable, though the existence of globally defined Hermitian structures is much more delicate (depending on the signature of spacetime). Our longer term goal is to use these ideas as background for a renewed attack on understanding the Newman--Janis trick~\cite{Newman:1965a, Newman:1965b, Giampieri, MSc, variations}.

%%------------------------------------------------------------------------------------------------------------------------------------------
\section{Strategy}
%%------------------------------------------------------------------------------------------------------------------------------------------

As much as possible we shall focus on the minimum technical machinery required to do useful physics:
\begin{itemize}
\itemsep-2pt
\item 
What is needed to set up basic spacetime structure?
\item 
What is needed to set up a sensible notion of causality?
\item
 What is needed to usefully extend the standard model of particle physics from Minkowski space to curved spacetime (ie, semiclassical gravity)?
\end{itemize}
The overall message to take from these considerations is that (even in the absence of full quantum gravity) physics is much more than just Lorentzian spacetime plus the Einstein equations. 
While the extra technical structure required to do useful physics is often implicit rather than explicit, we shall try to make the discussion as explicit and as closely coupled to direct observation as is reasonable.

%%------------------------------------------------------------------------------------------------------------------------------------------
\section{Preliminaries: Why manifolds?}
%%------------------------------------------------------------------------------------------------------------------------------------------

We first have to decide why we want to model spacetime by a \emph{manifold} based on $\R^4$.
There are actually three separate  physics issues here --- the choice of locally Euclidean space (based on $\R^4$), the Hausdorff condition, and the more highly technical second-countability/paracompactness condition.

--- Regarding the physical justification for the use of locally Euclidean space, once one thinks carefully about it, all physical measurements are based on ratios of (observed quantity)/(laboratory standard), and a good case can be made that physical measurements correspond to rational numbers, $\Q$, and finite-accuracy rational numbers at that~\cite{best}. Going to the reals $\R$, or the complex numbers $\C$, requires a step beyond the direct empirical evidence. 
Now pragmatically, that is a very useful step, since the reals $\R$, or the complex numbers $\C$, underly our ability to integrate and differentiate, and so set up the theory of ODEs and PDEs~\cite{best}.
This is pragmatically a very useful mathematical framework, so we should keep it as long as it is not in direct conflict with empirical reality. 
(There have been repeated suggestions that one should replace  $\R^4$ by something else once one gets to the Planck scale; but in the absence of direct empirical evidence to give us guidance, it is simply safer to stay with $\R^4$.)

--- The Hausdorff condition prevents ``branching'' or ``point-doubling'' topologies, (such as ``train track'' topologies or the ``bug-eyed line''). 
This is essential to setting up sensible existence and uniqueness theorems for ODEs and PDEs~\cite{math464}.
Again this is very closely tied to the utterly pragmatic necessity for setting up a sufficiently powerful mathematical ODE/PDE formalism to be able to do useful physics~\cite{best}.

---  In defining a manifold one typically adds yet another technical condition, that of second countability (or sometimes paracompactness). 
The physical import of this condition is not at all clear, and for physicists it is much better to replace this with the ``countable chart condition''; that there exists at least one countable atlas, (with at most a countable collection of charts), that covers the entire spacetime. 
That, (within the context of Hausdorff locally Euclidean spaces), these three conditions are equivalent is discussed in~\cite{Gauld1, Gauld2,  math464}. In fact Gauld has shown that within this context these three conditions are in turn equivalent to some 120-odd related technical conditions~\cite{Gauld1, Gauld2},  with one of the most central of the equivalencies being to the notion of  ``metrizability''.  
Unfortunately this is ``metrizability'' in the sense of being able to construct a Euclidean-signature metric, which is not of direct interest to physicists. ``Lorentzian metrizability'' requires Euclidean ``metrizability'' \emph{plus} the existence of an everywhere non-zero direction field (equivalent to a non-zero vector field up to sign)~\cite{centenary}, and a causally well behaved Lorentzian metric adds yet more conditions. 

For the time being let us simply assert this: The physicist's notion of a manifold requires locally Euclidean space, Hausdorff topology, and at most a countable number of coordinate patches.  

%%------------------------------------------------------------------------------------------------------------------------------------------
\section{Locally Minkowski structure: Causal structure}
%%------------------------------------------------------------------------------------------------------------------------------------------

Perhaps the key feature of general relativity is local flatness --- the locally Minkowski structure of spacetime encoded in the Einstein equivalence principle. While this to some extent correctly focusses attention on the spacetime metric, this is often done at the expense of other features.   
For instance, in addition to the light-cones themselves, defined ultra-locally at individual points in the spacetime, in any quasi-local finite region currently accessible to experiment/observation we need to be able to designate one of these light cones as ``future'' and the other as ``past''~\cite{Hawking-Ellis, Sachs-Wu}. In the absence of experimental/observational evidence to the contrary, it then makes sense to adopt this as a global requirement --- that everywhere in spacetime the light cones can be consistently divided into ``future'' and ``past''. 
This is the demand of ``time orientability'', note that it is a global constraint on the spacetime manifold, and that it is \emph{contingent} on the experimental/observational situation. (As is the Einstein equivalence principle itself.) 

Instead of working directly with the light cones themselves, time orientability is often rephrased in terms of the existence of a globally defined everywhere nonzero timelike unit vector field $V^a$. (So that $g(V,V)=-1$.) Time-orientability is the first of many  \emph{contingent} global properties it makes sense to demand. In fact, time orientability implies (Euclidean) metrizability --- by considering the (unphysical) Euclidean metric $g_E = g + 2\; V\otimes V$~\cite{centenary, Hawking-Ellis, Sachs-Wu}. (That is, $(g_E)_{ab} = g_{ab}+2V_a V_b$.) 

When put into a proper physical framework, Einstein gravity is more than just the equivalence principle plus the field equations. (If one \emph{only} uses the equivalence principle plus the field equations then the GIGO principle comes to the fore --- garbage in garbage out --- there are many \emph{formal} solutions to the equivalence principle plus the field equations that are physical nonsense~\cite{chronology}.)

Even within the framework of classical general relativity more can be done. For instance the existence of a Hamiltonian formalism (the ADM formalism) adds extra structure~\cite{toolkit}.  That we want a Hamiltonian structure to exist for general relativity is a consequence of the fact that general relativity is already known to have a Lagrangian structure (the Einstein--Hilbert Lagrangian, possibly supplemented by the Gibbons--Hawking surface term), together with the meta-theorem that Lagrangians can always be converted to Hamiltonians and vice-versa. Quasi-locally, on each finite spacetime region $U_{(\alpha)}$  the ADM formalism demands that spacetime be foliated by spacelike hypersurfaces, 
implying the (patch-wise) existence of some quasi-local time coordinate $\tau_{(\alpha)}$ such that $\nabla\tau_{(\alpha)}$ is nonzero and timelike throughout the region $U_{(\alpha)}$. This is a quasi-local patchwise version of ``stable causality''.  To bootstrap this to a global notion of stable causality can be justified in several ways: First,  in the absence of experimental/observational evidence to the contrary, there is no obstruction to adopting this as a global requirement. Second, more usual (non-GR) Hamiltonian structures are globally defined on some suitable phase space, so one should at least try to do the same in GR.  
In short, and \emph{contingent} upon the experimental/observational situation, it makes sense to demand global stable causality --- the existence of a globally defined scalar $\tau$ such that $\nabla \tau$ is everywhere non-zero and timelike. Note this automatically implies global time-orientability.

The next step is to demand hyperbolicity~\cite{Hawking-Ellis, Sachs-Wu, chronology, toolkit, Joshi, Penrose}. Almost everything we know about physics is intimately tied up with the existence of hyperbolic second-order differential equations, so quasi-locally, on each finite spacetime region $U_{(\alpha)}$  it is desirable to impose ``quasi-local hyperbolicity''. Again,  in the absence of direct experimental/observational evidence to the contrary, there is no obstruction to adopting this as a global requirement, and so it is common to demand the global hyperbolicity of spacetime. 
For all practical purposes this is the demand that $\mathcal{M} = \mathcal{R} \times \Sigma$, the demand that topologically spacetime factorizes into (time)$\times$(space)~\cite{Hawking-Ellis, Sachs-Wu}. Note that global hyperbolicity implies global stable causality which in turn implies global time-orientability.

This is about as far as one can get based on purely classical ideas. We shall now turn first to basic quantum mechanix, and then to curved-space QFT incorporating ideas from the standard model of particle physics. 

%%------------------------------------------------------------------------------------------------------------------------------------------
\section{Quantum mechanix}
%%------------------------------------------------------------------------------------------------------------------------------------------

Basic quantum mechanix (even before we concern ourselves with curved-spacetime QFT) already contains Hamiltonian operators, plus the parity $P$ and time-reversal $T$ operators. Even the simplest versions of quantum mechanix need a Hamiltonian to be globally defined on space (for at least a finite amount of time),  so one should at least try to do the same in GR. This is an additional reason for wanting to apply the ADM formalism globally, and so an additional reason for demanding global stable causality.  

The empirical success of the time-reversal concept even in non-relativistic quantum mechanix makes it clear that we can physically distinguish ``future'' from ``past'', so one should at least try to do the same in GR. This is an additional reason for demanding global time-orientabilty.  

Furthermore the empirical success of the parity concept even in non-relativistic quantum mechanix makes it clear that we can physically distinguish ``left-handed'' from ``right-handed'' coordinate systems. Quasi-locally at least space should be orientable. Combined with the previously discussed notion of quasi-local time orientability, this implies that spacetime should be quasi-locally time-orientable, space-orientable, and spacetime-orientable. (Quasi-local time-orientability requires the existence of a non-zero timelike vector field; quasi-local space-orientability requires the existence of a triad of linearly independent spacelike vector fields; combining the two quasi-local spacetime orientability demands the existence of a tetrad of linearly independent vector fields.)

Again,  in the absence of direct experimental/observational evidence to the contrary, there is no obstruction to adopting this as a global requirement, and so it is common to demand global time-orientability, global space-orientability, and global spacetime-orientability. Now global time-orientability requires the existence of a globally defined non-zero timelike vector field; global space-orientability requires the global existence of a triad of linearly independent spacelike vector fields; combining the two global spacetime-orientability demands the global existence of a tetrad of linearly independent vector fields. Note that the global existence of tetrads/vierbeins has already snuck up on us based ``merely'' on time-orientability and parity, even before explicitly introducing fermions. We will have much more to say about this in the next section where we discuss chiral fermions in the standard model of particle physics. 

%%------------------------------------------------------------------------------------------------------------------------------------------
\section{Standard model of particle physic: Chiral fermions}
%%------------------------------------------------------------------------------------------------------------------------------------------

The key features of the standard model of particle physics we shall be interested in are the existence of  the  $C$, $P$, and $T$ operators, and the existence of 
chiral fermions. The $P$ and $T$ operators already make sense at the level of quantum mechanics, and were discussed above. The $C$ operator seems to have less direct relevance to spacetime \emph{per~se}. The existence of chiral fermions is much more significant. 

Chiral fermions require two key technical conditions to be satisfied: the existence of chiral (two-component) spinor fields, and the existence of the  chiral (two-component) Dirac equation $\slashed{D} \psi = 0$. While these concepts are initially constructed in flat Minkowski space, the fact that empirically and pragmatically they continue to make sense in weak gravitational fields indicates that they have at least quasi-local significance in general relativity~\cite{Joshi, Penrose, Penrose-Rindler, O'Donnell, Parker-Toms, Chandrasekhar}. 

As is by now usual,  in the absence of direct experimental/observational evidence to the contrary, there is no obstruction to adopting the existence of 2-spinors as a global requirement, and the only global \emph{topological} obstruction to doing so is the 2$^{nd}$ Stiefel--Whitney class. So it is usual to demand that the  2$^{nd}$ Stiefel--Whitney class vanish. 
There is a highly technical route from the vanishing of the 2$^{nd}$ Stiefel--Whitney to the existence of global tetrads~\cite{Lee1, Lee2, Geroch}, but there is also a more direct physical route.
 
 One additionally needs to be able to define the (chiral) Dirac equation $\slashed{D} \psi = 0$.  Unwrapping this, we see $\gamma^a(x) \partial_a \psi = 0$, so one needs to somehow be able to define local Dirac matrices $\gamma^a(x)$ such that $\{\gamma^a(x),\gamma^b(x)\} = 2 g^{ab}(x)$. 
The most common way of implementing this is by taking the local Dirac matrices $\gamma^a(x)$ to be linear combinations of the usual Minkowski space Dirac matrices as follows
\begin{equation}
\gamma^a(x) = e^a{}_A(x)  \;\gamma^A; \qquad \{\gamma^A,\gamma^B\} = 2 \eta^{AB}; \qquad  
e^a{}_A(x)  \;e^b{}_B(x)\; \eta^{AB} = g^{ab}(x).
\end{equation}
The $e^a{}_A(x)$ are just seen to be an ordinary tetrad. (The index $a$ is a coordinate index while the label $A$ simply labels which of the four vectors in the tetrad one is dealing with.) For the tetrad itself one could equivalently write
\begin{equation}
e_a{}^A(x)  \;e_b{}^B(x)\; \eta_{AB} = g_{ab}(x).
\end{equation}
Thus the empirical existence of fermions \emph{forces} us to adopt a variant of the tetrad/ vierbein formalism. 
Even if one were to attempt to eschew tetrads, (see for instance Weldon~\cite{Weldon:2000}), the local Dirac matrices $\gamma^a(x)$ would still be needed, and these local Dirac matrices would implicitly be hiding the tetrad in their formulation. 

Direct empirical evidence only guarantees quasi-local patch-wise existence of these tetrads, but as is by now usual,  in the absence of direct experimental/observational evidence to the contrary, there is no obstruction to adopting the existence of tetrads as a global requirement. This agrees with our previous arguments based on quantum mechanix and the $P$ and $T$ operators, but now there is a much more focussed and mission-critical issue at play: One simply cannot set up curved-space fermions without at least quasi-local tetrads, and the global existence of a tetrad field is at the very least highly desirable. In particular this strongly suggests (despite what classical general relativists might feel) that the metric might not be the fundamental physical quantity --- a very good case can be made for the tetrad being more fundamental than the metric.

%%------------------------------------------------------------------------------------------------------------------------------------------
\section{Minkowski-space tetrads/vierbeins}
%%------------------------------------------------------------------------------------------------------------------------------------------

Above we have argued for the existence of a globally defined tetrad of linearly independent vector fields. (Consequently the determinant $\det(e_a{}^A)\neq 0$, and so it can always be chosen to be of fixed sign so that the spacetime is spacetime-orientable.) Naively one might think that the tetrad contains one timelike and three spacelike vector fields, but the situation is much trickier. 
It is (perhaps unexpectedly) possible to arrange all four vectors in tetrad to be timelike, or all four to be spacelike, or all four to be null. Comments along these lines date back (at least) to Synge in the 1950's~\cite{Synge}. 

Working temporarily in $-+++$ signature Minkowski space consider the tetrad
\begin{equation}
\{  (+2;0,0,0); \quad (+2;+1,0,0); \quad (+2;0,+1,0); \quad (+2;0,0,+1) \}.
\end{equation}
All four vectors are timelike, but the tetrad is linearly independent and spans the entire Minkowski space. 
(You can adapt this argument to construct 4 timelike coordinates [non-orthogonal] covering all of Minkowski space.)

Now consider the Minkowski space tetrad
\begin{equation}
\{  (+1;-2,0,0); \quad (+1;+2,0,0); \quad (+1;0,+2,0); \quad (+1;0,0,+2) \}.
\end{equation}
All four vectors are spacelike, but the tetrad is linearly independent and spans the entire Minkowski space. 
(You can adapt this argument to construct 4 spacelike coordinates [non-orthogonal] covering all of Minkowski space.)
Now consider the Minkowski space tetrad
\begin{equation}
\{  (+1;-1,0,0); \quad (+1;+1,0,0); \quad (+1;0,+1,0); \quad (+1;0,0,+1) \}. 
\end{equation}
All four vectors are null (lightlike), but the tetrad is linearly independent and spans the entire Minkowski space. 
(You can adapt this argument to construct 4 null coordinates [non-orthogonal] covering all of Minkowski space.)

The point is that there is a lot of freedom in setting up tetrads, while choosing one timelike and three spacelike vectors is a particularly common choice, it is by no means the only way to set things up. Another particularly common choice in Minkowski space is a \emph{complex} null tetrad
\begin{equation}
\{  (+1;-1,0,0); \quad (+1;+1,0,0); \quad (0;0,+1,+i); \quad (0;0,1,-i) \}. 
\end{equation}
Note that the spacetime coordinates are kept real, as is the metric $\eta_{ab}$. It is only that \emph{some} of the components of the tetrad vectors $e^a{}_A$ are allowed to go complex, the underlying manifold and metric are both still real. 

%%------------------------------------------------------------------------------------------------------------------------------------------
\section{Curved spacetime tetrads/vierbeins}
%%------------------------------------------------------------------------------------------------------------------------------------------

All of these considerations carry over to curved spacetime where one particularly canonical choice for the tetrad structure is the (time)+(space) split
\begin{equation}
\{ e^a{}_A \} = \{ V^a;  S^a{}_k \};  \qquad g(V,V)=-1; \qquad g(V,S_i)=0; 
\qquad g(S_j,S_k) = \delta_{jk}.
\end{equation}
(This particular tetrad has the virtue of making time-orientability manifest.)  
For this (space)+(time) orthonormal tetrad we have
\begin{equation}
g(e_A,e_B)  = g_{ab} \;  e^a{}_A \; e^b{}_B = \eta_{AB} =
\left[  \begin{array}{cc|cc} -1 & 0 & 0 & 0\\ 0&+1&0&0\\\hline0&0&+1&0\\0&0&0&+1\end{array}\right]_{AB}.
\end{equation}
%-----------------------------------------------------------------------------

A second canonical choice is the complex null tetrad
\begin{equation}
\{ e^a{}_A \} = \{ \ell^a,  n^a, m^a, \bar m^a \};  \qquad g(\ell,\ell)=g(n,n)=g(m,m)= g(\bar m,\bar m) =0.
\end{equation}
Indeed for the complex null tetrad it is common to use the full freedom of  taking arbitrary linear combinations of the tetrad fields to (without loss of generality) choose the \emph{orthonormal} complex null tetrad
\begin{equation}
g(e_A,e_B)  = g_{ab} \;  e^a{}_A \; e^b{}_B = \eta_{AB} = 
\left[  \begin{array}{cc|cc} 0 & -1 & 0 & 0\\ -1&0&0&0\\\hline0&0&0&+1\\0&0&+1&0\end{array}\right]_{AB}.
\end{equation}
Perhaps unexpectedly
\begin{equation}
\{ e^a{}_A \} = \{ \ell^a,  n^a, m^a, \bar m^a \};  \qquad \implies \qquad \{ e_a{}^A \} = \{ -n_a,  -\ell_a, \bar m_a,  m_a \}.
\end{equation}
This is equivalent to choosing~\footnote{We prefer to work in $-+++$ signature where $\ell^an_a = n^a \ell_a = -1$.}
\begin{equation}
g_{ab} =  -(\ell_a n_b + n_a \ell_b) + m_a \bar m_b + \bar m_a m_b,
\end{equation}
or equivalently
\begin{equation}
g^{ab} =  -(\ell^a n^b + n^a \ell^b) + m^a \bar m^b + \bar m^a m^b.
\end{equation}
Several points should be made here:
\begin{enumerate}
\item
For a fixed metric, the tetrad is not unique, being specified only up to a local Lorentz transformation~\cite{Chandrasekhar}. 
\item 
From the above we see that the existence of a global complex null tetrad is not a significant extra technical assumption on the spacetime; it essentially comes along for free as soon as one realises that we simply will have to confront curved space fermions as part of our physics. 
\item
The ``complex'' in ``complex null tetrad'' can perhaps mislead the unwary; the underlying spacetime is perfectly real, (the coordinates are real numbers), it is only \emph{some} vector components, (used at intermediate steps of the calculation), that are allowed to go complex. The spacetime metric, and in fact all of the physics, is again perfectly real. (We belabour this point now with a view to future planned work on the Newman--Janis ansatz~\cite{variations}.) 
\item
The existence of a global tetrad (whether real or complex null) implies that the manifold is \emph{parallelizable}. Again, this is not a significant extra assumption on the spacetime; it essentially comes along for free as soon as one realises that we simply will have to confront curved spacetime fermions as part of our physics. 
\item
Because the manifold is  \emph{parallelizable} we can (\emph{if desired}) construct a Riemann-flat Weitzenbock connexion. Specifically take~\footnote{The Weitzenbock connexion is only unique if the tetrad is specified. If it is only the metric that is specified, then the Weitzenbock connexion is defined only up to a Local lorentz transformation.}
\begin{equation}
[\Gamma_\mathrm{Weitzenbock}]^a{}_{bc} = e^a{}_A\; \partial_c e_b{}^A = - e_b{}^B\; \partial_c e^a{}_B.
\end{equation}
It is then an easy exercise to check that the Christoffel symbol decomposes as
\begin{equation}
\Gamma^a{}_{bc} =   g^{am} \;\eta_{AB} \; \left\{ e_{m}{}^{A}  e_{[b,c]}{}^B  + e_{[m,b]}{}^A e_c{}^B + e_{[m,c]}{}^A e_b{}^B{} \right\}     +[\Gamma_\mathrm{Weitzenbock}]^a{}_{bc}.
\end{equation}
Whether one \emph{wants} to do so is a separate physical question, the fact that one \emph{can} do so essentially comes along for free as soon as one realises that we simply will have to confront curved spacetime fermions as part of our physics. 
\end{enumerate}
Up to this stage we have tried to stay very close to empirical reality, using only the most basic and non-controversial aspects of the curved-spacetime extension of standard model of particle physics. There are additional structures, based on complex manifold theory, that one might wish to consider --- but these are much more speculative and much less directly tied to empirical reality. 

%%------------------------------------------------------------------------------------------------------------------------------------------
\section{Almost complex structures}
%%------------------------------------------------------------------------------------------------------------------------------------------

Up to this stage, we have adduced good reasons for demanding the existence of a global orthonormal complex null tetrad, $e_a{}^A$, with some complex components, but defined on a real spacetime manifold.  Let us now further define
\begin{equation}
J^A{}_B= 
\left[  \begin{array}{cc|cc} 0& +1 & 0 & 0\\ -1&0&0&0\\\hline0&0&+i&0\\0&0&0&-i\end{array}\right]^A_{~~B}\;; \qquad
\hbox{so that} \qquad  J^A{}_B \, J^B{}_C= - \delta^A{}_C. 
\end{equation}
This is equivalent to defining
\begin{equation}
J^a{}_b  = e^a{}_A \; J^A{}_B\; e_b{}^B =  -(\ell^a \ell_b - n^a n_b) + i(m^a \bar m_b - \bar m^a m_b); \qquad  
J^a{}_b \, J^b{}_c = - \delta^a{}_c.
\end{equation}
This is a globally defined  ``almost complex structure'' in the usual (mathematical) sense. 
See for instance Vandoren~\cite{Vandoren} and Flaherty~\cite{Flaherty:thesis, Flaherty:book, Flaherty:pla}.
This almost complex structure is not unique, since it depends on the particular tetrad one chooses to represent the metric. 
Note that the tensor $  i(m^a \bar m_b - \bar m^a m_b)$ is pure real, (so that $J^a{}_b$ is real, even though $J^A{}_B$ has some imaginary components). Note $J^A{}_A = 0 = J^a{}_a$. 
Now consider
\begin{equation}
J^{AB}=  J^A{}_C \; \eta^{CB} =  
\left[  \begin{array}{cc|cc} -1& 0 & 0 & 0\\ 0&+1&0&0\\\hline0&0&0&+i\\0&0&-i&0\end{array}\right]^{AB}\;;
\quad
J^{ab}  = -(\ell^a \ell^b - n^a n^b) + i(m^a \bar m^b - \bar m^a m^b).
\end{equation}
Then $J^{AB}$ is Hermitian, while $J^{ab}$ is real symmetric (and \emph{not} equal to $g^{ab}$).
Up to this stage everything is very standard. 
However
\begin{equation}
J^A{}_B \, J^C{}_D \, \eta^{BD} \neq  \eta^{AC}; \qquad \implies \qquad J^a{}_b\, J^c{}_d\, g^{bd} \neq g^{ac}.
\end{equation}
So with this almost complex structure the spacetime metric is \emph{not} ``Hermitian'' in the sense usually used in complex manifold theory.  In fact
\begin{eqnarray}
J^a{}_b\, J^c{}_d\, g^{bd}  &=& J^a{}_b\, J^c{}_d  \left\{ -(\ell^a n^b + n^a \ell^b) + m^a \bar m^b + \bar m^a m^b \right\}
\nonumber\\
&=&  +(\ell^a n^b + n^a \ell^b) + m^a \bar m^b + \bar m^a m^b \neq g^{ac}.
\end{eqnarray}
The unwanted sign flip in the $\ell n$ sector can eventually be traced back to Lorentzian signature of the spacetime manifold~\cite{Flaherty:thesis, Flaherty:book, Flaherty:pla}.
Perhaps the key message to take from this is that one cannot \emph{blindly} copy over results from complex manifold theory. 
This is one area where the distinction between Lorentzian signature and Euclidean signature is important, and some care and delicacy must be used when attempting to use complex manifold technology in Lorentzian spacetime physics. 

%%------------------------------------------------------------------------------------------------------------------------------------------
\section{Modified almost complex structures}
%%------------------------------------------------------------------------------------------------------------------------------------------

One attempt at more closely merging complex manifold theory with spacetime physics was through the adoption of Flaherty's modified almost complex structure~\cite{Flaherty:thesis, Flaherty:book, Flaherty:pla}. Let us define
\begin{equation}
J^A{}_B= 
\left[  \begin{array}{cc|cc} i & 0 & 0 & 0\\ 0&-i&0&0\\\hline0&0&i&0\\0&0&0&-i\end{array}\right]^A_{~~B}\;; \qquad
\hbox{so that} \qquad  J^A{}_B \, J^B{}_C= - \delta^A{}_C. 
\end{equation}
This is equivalent to defining
\begin{equation}
J^a{}_b  = e^a{}_A \; J^A{}_B\; e_b{}^B =  -i (\ell^a n_b - n^a \ell_b) + i(m^a \bar m_b - \bar m^a m_b); \qquad  
J^a{}_b \, J^b{}_c = - \delta^a{}_c.
\end{equation}
This is certainly globally defined, but it is not an an ``almost complex structure'' in the usual mathematical sense. 
This modified almost complex structure is not unique, since it depends on the particular tetrad one chooses to represent the metric. 
Note that the tensor $ i (\ell^a n_b - n^a \ell_b) $ is pure imaginary while the tensor $  i(m^a \bar m_b - \bar m^a m_b)$ is pure real. 
Some of the components of the tensor $J^a{}_b$ are unavoidably complex, and this ultimately tracks back to the Lorentzian signature of spacetime~\cite{Flaherty:thesis, Flaherty:book, Flaherty:pla}. Now consider
\begin{equation}
J^{AB}=  J^A{}_C \; \eta^{CB} =  
\left[  \begin{array}{cc|cc} 0& -i & 0 & 0\\ +i&0&0&0\\\hline0&0&0&+i\\0&0&-i&0\end{array}\right]^{AB}\;;
\quad
J^{ab}  = -i(\ell^a n^b - n^a \ell^b) + i(m^a \bar m^b - \bar m^a m^b).
\end{equation}
Then $J^{AB}$ is Hermitian, while $i(m^a \bar m^b - \bar m^a m^b)$ is real symmetric, and $-i(\ell^a n^b - n^a \ell^b)$ is imaginary and anti-symmetric, so  $J^{ab}$ is Hermitian. 
Note that we now have
\begin{equation}
J^A{}_B \, J^C{}_D \, \eta^{BD} =  \eta^{AC}; \qquad \implies \qquad J^a{}_b\, J^c{}_d\, g^{bd} = g^{ac}.
\end{equation}
This is the closest one can get (in Lorentzian signature spacetime) to the notion of a ``Hermitian'' metric, but only for this nonstandard ``modified'' notion of almost complex structure. Unfortunately this formalism exhibits other drawbacks, and has not really caught on in the general relativity community. (In particular, if one manages to set up complex coordinates,  then  in this formalism the complex coordinate $\bar z$ is not the complex conjugate of the complex coordinate $z$, see~\cite{Flaherty:thesis, Flaherty:book, Flaherty:pla}. This is a less than desirable state of affairs.)

%%------------------------------------------------------------------------------------------------------------------------------------------
\section{Spacetime is not Hermitian}
%%------------------------------------------------------------------------------------------------------------------------------------------

For a deeper look at what went wrong when attempting to set up a Hermitian metric on spacetime, consider the considerably simpler (1+1) dimensional case. We know we can always adopt local coordinates such that
\begin{equation}
\d s^2 = h(t,x) \left\{ -\d t^2 +\d x^2 \right\}.
\end{equation}
Let us define $z=t+ix$ and $\bar z = t -ix$, then $t=(z+\bar z)/2$ and $x= (z-\bar z)/(2i)$, so that
\begin{equation}
\d s^2 = - {h(z,\bar z)\over2}  \left\{\d z^2 +\d \bar z^2 \right\}.
\end{equation}
This is certainly a real metric on a complex coordinate patch, but it is not what is normally referred to as Hermitian within the context of complex manifold theory. (A Hermitian metric would be of the form $\d s^2 = g_{z\bar z}(z,\bar z) \; \d z\; \d \bar z$,  but this is automatically of Euclidean signature,  and a Lorentzian metric can never be put in this form.)

This result extends to higher dimensionality: Consider  a $n$ complex dimensional Hermitian metric  $\d s^2 = g_{i\bar j}(z,\bar z) \; \d z^i\; \d \bar z^j$ and use local coordinate transformations to diagonalize the Hermitian matrix $g_{i\bar j}(z,\bar z)$.
Then the manifold is of some complex signature $(n_+,n_-)$, but in terms of real coordinates this is signature $(2n_+,2n_-)$, which is certainly never of Lorentzian form $(1,2n-1)$.  Oddly enough, with two (or more generally any an even number) of time dimensions this obstruction goes away, though that introduces another set of physical problems.

%%------------------------------------------------------------------------------------------------------------------------------------------
%%------------------------------------------------------------------------------------------------------------------------------------------
\section{Discussion}
%%------------------------------------------------------------------------------------------------------------------------------------------
%%------------------------------------------------------------------------------------------------------------------------------------------

In this article we have tried to carefully explain why Einstein gravity (when put in its proper physical context) is much more than just the quasi-local physics of the spacetime metric. Extra global structure (orientability, stable causality, hyperbolicity, tetrads, parallelizability) are not just optional extras, they are automatically introduced as soon as one tries to incorporate parts of the standard model of particle physics into curved-space QFT. In particular tetrads and parallelizability are essentially unavoidable as soon as one tries to  implement either $T$ and $P$ operators or general relativistic fermions. Even to this day, many general relativity textbooks are unclear on this point, and this article has been written with a view to aiding cross-communication between the sub-disciplines.

In contrast, imposing complex structure is a much more ambitious proposal which is much less tied to empirical observations. 
While any tetrad leads naturally to an almost complex structure, this almost complex structure is tetrad dependent and (in terms of the metric) not unique.  More troublingly, in Lorentzian signature the almost complex structure is not compatible with any almost Hermitian structure, and attempts at side-stepping this issue seem to generate more problems than they solve. 

We had originally hoped to be able to use this physical framework for a renewed attack on understanding the physics (if any) underlying the  Newman--Janis ansatz, and in this regard the situation is very mixed. Yes, it is clear from the above that the use of complex null tetrads (intrinsic to the original Newman--Janis ansatz) is physically well motivated. On the other hand, the complex transformations inherent to the 
Newman--Janis ansatz are still somewhat mysterious. We plan to address these issues more fully in future work. 

%------------------------------------------------------------------------------------------------------------------------------------------
%------------------------------------------------------------------------------------------------------------------------------------------
\acknowledgments
%%------------------------------------------------------------------------------------------------------------------------------------------
%%------------------------------------------------------------------------------------------------------------------------------------------

This research was supported by the Marsden Fund, through a grant administered by the Royal Society of New Zealand. 
\\
DN was also supported by a Victoria University of Wellington MSc Scholarship.
%---------------------------------------------------------------------------------------------------------------------------------------------

%------------------------------------------------------------------------------------------------------------------------------------------
%------------------------------------------------------------------------------------------------------------------------------------------
\end{document}